\documentclass[aps,preprint,showpacs,superscriptaddress,groupedaddress]{revtex4}  

\usepackage{amssymb}
\usepackage{amsmath}
\usepackage{amssymb}
\usepackage{graphicx}
\usepackage{subfigure}
\usepackage{color}
\usepackage{mathrsfs} 
\usepackage{hyperref}
\usepackage{soul}

\usepackage[utf8]{inputenc}
\usepackage[english]{babel}

\begin{document}

\title{Explicit kinks in higher-order field theories}

\author{Vakhid A. Gani}
\email{vagani@mephi.ru}
\affiliation{Department of Mathematics, National Research Nuclear University MEPhI (Moscow Engineering Physics Institute),
Moscow, 115409 Russia}
\affiliation{Theory Department, Institute for Theoretical and Experimental Physics of National Research Centre ``Kurchatov Institute'', 
Moscow, 117218 Russia}

\author{Aliakbar Moradi Marjaneh}
\email{moradimarjaneh@gmail.com}
\affiliation{Department of Physics, Quchan Branch, Islamic Azad university, Quchan, Iran}

\author{Petr A. Blinov}
\email{petr.blinov.mipt@gmail.com}
\affiliation{Moscow Institute of Physics and Technology,
Dolgoprudny, Moscow Region, 141700 Russia}

\begin{abstract}

We study a field-theoretic model with an eighth-degree polynomial potential --- the $\varphi^8$ model. We show that for some certain ratios of constants of the potential, the problem of finding kink-type solutions in $(1+1)$-dimensional space-time reduces to solving algebraic equations. For two different ratios of the constants, which determine positions of the vacua, we obtained explicit formulas for kinks in all topological sectors. The properties of the obtained kinks are also studied --- their masses are calculated, and the excitation spectra which could be responsible for the appearance of resonance phenomena in kink-antikink scattering are found.

\end{abstract}



\maketitle

\section{Introduction}
\label{sec:introduction}

Topological solitons is an important class of solutions of field-theoretic models which are of great importance to high energy physics, cosmology, and condensed matter \cite{Rajaraman.book.1982,Vilenkin.book.2000,Manton.book.2004,Vachaspati.book.2006,Kevrekidis.book.2019}. In this context, models with polynomial potentials, in turn, are widely used. Apart from applications in high energy physics theory, such models are used to simulate spontaneous symmetry breaking in the Ginzburg-Landau model of superconductivity \cite{Landau.ZhETF.1937,Ginzburg.ZhETF.1950}, (consecutive) phase transitions in materials \cite{Khare.PRE.2014,Gufan.DAN.1978}, field evolution in the early Universe \cite{Gani.JCAP.2018} etc., see also \cite{Bishop.PhysD.1980,Belova.UFN.1997,Kevrekidis.book.2019} for review.

The great progress has been made in the study of models with both polynomial and non-polynomial potentials. In particular, the {\it deformation procedure} has been developed \cite{Bazeia.PRD.2002,Bazeia.PRD.2004,Bazeia.PRD.2006} that allows one to find new models (potentials) and simultaneously their one-dimensional topologically nontrivial solutions --- {\it kinks} \cite[Chap.~5]{Manton.book.2004}. There are impressive advances in the study of kinks scattering and multi-kink interactions in such models as, e.g., $\varphi^4$ model \cite[Chaps.~1--11 and 13]{Kevrekidis.book.2019}, \cite{Kudryavtsev.JETPLett.1975,Campbell.PhysD.1983,Belova.PhysD.1988,Goodman.SIAM_JADS.2005,Moradi.CNSNS.2017,Dorey.JHEP.2017,Dorey.PLB.2018,Weigel.JPCS.2014,Weigel.PRD.2016}, $\varphi^6$ model \cite[Chap.~12]{Kevrekidis.book.2019}, \cite{Weigel.JPCS.2014,Weigel.PRD.2016,Hoseinmardy.IJMPA.2010,Dorey.PRL.2011,Gani.PRD.2014,Romanczukiewicz.PLB.2017,Weigel.PLB.2017,Weigel.AHEP.2017,Moradi.JHEP.2017,Demirkaya.JHEP.2017,Lima.JHEP.2019,Lohe.PRD.1979}, $\varphi^8$, $\varphi^{10}$, $\varphi^{12}$ models \cite{Khare.PRE.2014}, \cite[Chap.~12]{Kevrekidis.book.2019}, \cite{Lohe.PRD.1979,Gani.JHEP.2015,Christov.PRD.2019,Belendryasova.CNSNS.2019,Christov.PRL.2019}, various modifications of the sine-Gordon model \cite{Peyrard.PhysD.1983.msG,Campbell.PhysD.1986.dsG,Gani.PRE.1999,Bazeia.EPJC.2011,Moradi.EPJB.2018,Gani.EPJC.2018,Belendryasova.JPCS.2019,Gani.EPJC.2019}, as well as in models with more exotic dynamics \cite{Bazeia.EPJC.2013,Zhong.JHEP.2014,Mendonca.JHEP.2015,Simas.JHEP.2016,Bazeia.EPJC.2018,Bazeia.PLB.2019,Zhong.PLB.2018,Bazeia.IJMPA.2019,Zhong.JHEP.2020} and multi-field models \cite{Alonso-Izquierdo.AHEP.2013,Katsura.PRD.2014,Alonso-Izquierdo.PRD.2018,Alonso-Izquierdo.PS.2019,Alonso-Izquierdo.CNSNS.2019,Gani.JHEP.2016}.

Recently, models with potentials in the form of polynomials of eighth degree and higher are of growing interest \cite{Khare.PRE.2014,Lohe.PRD.1979,Gani.JHEP.2015,Christov.PRD.2019,Belendryasova.CNSNS.2019,Christov.PRL.2019,Manton.JPA.2019}. In particular, the excitation spectra of the $\varphi^8$ kinks with exponential asymptotics, as well as resonance phenomena in the scattering of such kinks at low energies, were studied \cite{Gani.JHEP.2015}. As a separate branch of study, one can emphasize investigation of properties of kinks with power-law asymptotics. Such kinks are topological solutions of, e.g., the $\varphi^8$, $\varphi^{10}$, or $\varphi^{12}$ model with particular form of potential. More specifically, the potential must have a minimum, which is a zero of the fourth or higher order. Then the corresponding kink has a power-law asymptotic behavior at that spatial infinity at which the field approaches the aforementioned minimum, see, e.g., \cite[Sec.~II.A]{Christov.PRD.2019}. Due to the presence of power-law tails, kinks acquire new properties. In particular, a kink and an antikink (or a kink and a kink) placed at a certain distance from each other interact much more strongly than in the case of exponential asymptotics. This phenomenon is called {\it long-range interaction of kinks with power-law tails} \cite{Christov.PRD.2019,Belendryasova.CNSNS.2019,Christov.PRL.2019,Manton.JPA.2019}. In Ref.~\cite{Belendryasova.CNSNS.2019} scattering of the $\varphi^8$ kinks with power-law asymptotic behavior has been studied numerically. Besides, it was shown that resonance phenomena in the kink-antikink collisions could be a consequence of the presence of the vibrational modes of the ``kink+antikink'' system as a whole.

Recent works \cite{Christov.PRD.2019} and \cite{Christov.PRL.2019} continued the study of interactions of kinks with power-law asymptotics. It was demonstrated that the presence of the power-law tails entails long-range interaction which, in turn, requires a special approach to constructing of the initial conditions for the numerical simulations of the kink-antikink and kink-kink collisions. The problem is that the conventional initial conditions which were used in the case of exponential asymptotics, being applied to the kinks with power-law tails lead to appearance of significant disturbances due to radiation. This, in turn, creates the illusion of repulsion between kink and antikink \cite{Belendryasova.CNSNS.2019}. Several methods for ``distilling'' the initial configurations into suitable ans\"atze were proposed, it was shown how these approaches capture the attractive nature of interactions between the kink and antikink in the presence of long-range interaction \cite{Christov.PRD.2019}. The general results on the interactions of kinks with power-law asymptotics in $\varphi^{2n+4}$ models for $n\ge 2$ have been obtained in \cite{Christov.PRL.2019}. It was found that the interaction between kink and antikink is generically attractive, while the interaction between two kinks is generically repulsive. The force of interaction falls off with distance as its $2n/(n-1)$-th power. The obtained analytic estimation is in good agreement with the results of numerical simulations for $n=2$ (the $\varphi^8$ model), $n=3$ (the $\varphi^{10}$ model) and $n=4$ (the $\varphi^{12}$ model). It is worth to mention also Refs.~\cite{Manton.JPA.2019, Khare.JPA.2019}, where various properties of field-theoretic models with high-degree polynomial potentials are also considered.

Despite some certainly interesting attempts to obtain explicit expressions for kinks of models with polynomial potentials \cite{Bazeia.PRD.2006,Bazeia.AP.2018} with very specific set of vacua, kinks of the $\varphi^8$ model and of higher degree models so far could mainly be obtained in the implicit form. The purpose of this paper is to show that in some more general cases (for some relations between model parameters) it is possible to obtain explicit formulas for kinks. We will consider the example of the $\varphi^8$ model with a potential of a certain type, which will be described below.

This our paper is organized as follows. In Sec.~\ref{sec:model} we briefly describe the $\varphi^8$ model. In Sec.~\ref{sec:kinks} we show how the explicit formulas for kinks can be obtained for some particular model parameters. Section \ref{sec:properties} presents some properties of the obtained kink solutions. Finally, we conclude in Sec.~\ref{sec:conclusion}.

\section{The $\varphi^8$ model}
\label{sec:model}

\begin{figure}
    \centering
    \includegraphics[width=0.6
 \textwidth]{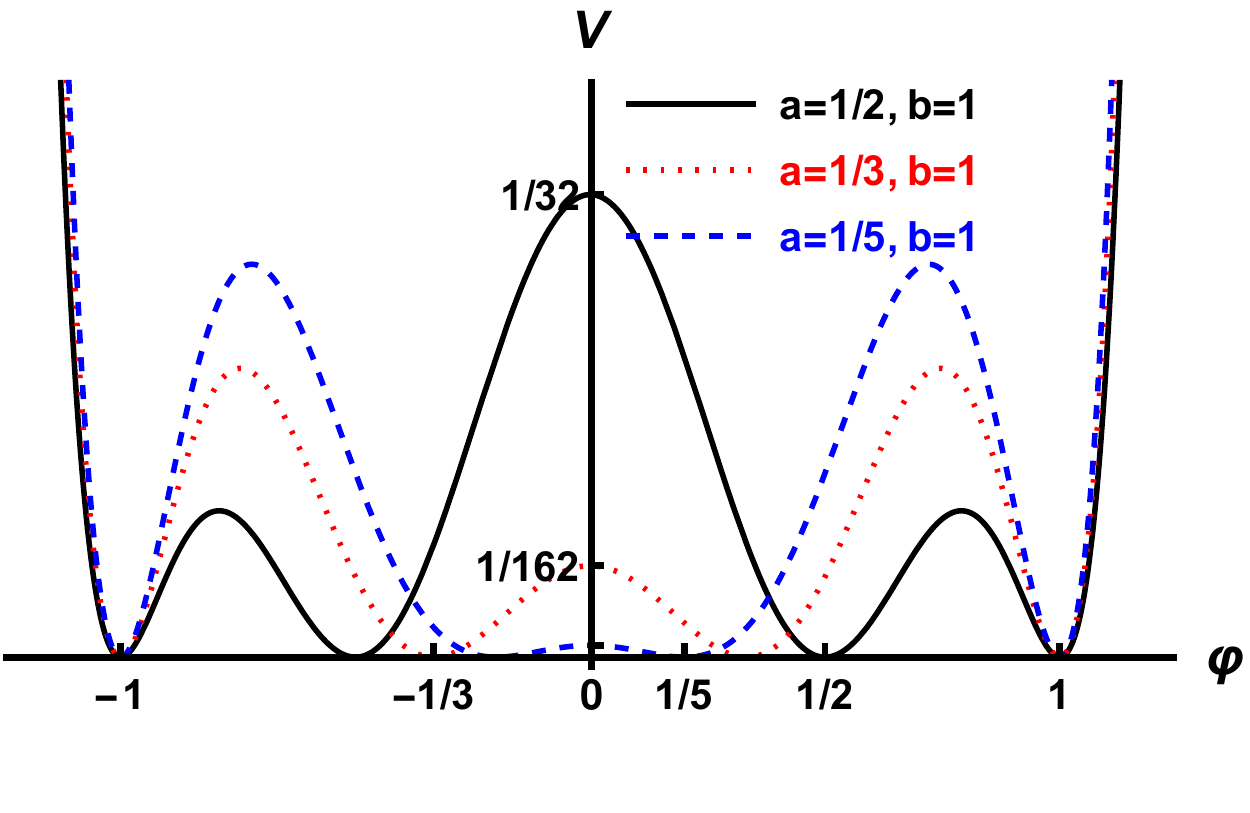}
 \\
    \caption{The potential \eqref{eq:potential} of the $\varphi^8$ model for $b=1$ and $a=\displaystyle\frac{1}{2}$, $\displaystyle\frac{1}{3}$, $\displaystyle\frac{1}{5}$.} 
    \label{fig:Potentials}
\end{figure}

Consider a field-theoretic model in $(1+1)$-dimensional space-time with a real scalar field $\varphi(x,t)$. Assume that the dynamics of the system is determined by the Lagrangian
\begin{equation}\label{eq:Largangian}
	\mathscr{L} = \frac{1}{2} \left( \frac{\partial\varphi}{\partial t} \right)^2 - \frac{1}{2} \left( \frac{\partial\varphi}{\partial x} \right)^2 - V(\varphi).
\end{equation}
For the topological kinks to exist, it is necessary that the potential $V(\varphi)$ be a (usually non-negative) function of the field $\varphi$ that has two or more degenerate minima. The model considered by us is described by the potential in the form of eighth degree polynomial:
\begin{equation}\label{eq:potential}
    V(\varphi) = \frac{1}{2}\left(\varphi^2-a^2\right)^2\left(\varphi^2-b^2\right)^2,
\end{equation}
where $a$ and $b$ are constants, $0<a<b$, see Fig.~\ref{fig:Potentials}.

The energy functional for the Lagrangian \eqref{eq:Largangian} is
\begin{equation}\label{eq:energy}
	E[\varphi] = \int_{-\infty}^{\infty}\left[\frac{1}{2} \left( \frac{\partial\varphi}{\partial t} \right)^2+\frac{1}{2} \left( \frac{\partial\varphi}{\partial x} \right) ^2+V(\varphi)\right]dx,
\end{equation}
which in the static case becomes
\begin{equation}\label{eq:stenergy}
	E_{\rm static}^{}[\varphi] = \int_{-\infty}^{\infty}\left[\frac{1}{2} \left( \frac{d\varphi}{d x} \right) ^2+V(\varphi)\right]dx.
\end{equation}
From the Lagrangian \eqref{eq:Largangian} one can obtain the equation of motion for the field $\varphi(x,t)$:
\begin{equation}\label{eq:eom}
	\frac{\partial^2\varphi}{\partial t^2}-\frac{\partial^2\varphi}{\partial x^2}+\frac{dV}{d\varphi}=0,
\end{equation}
which in the static case $\varphi=\varphi(x)$ takes the form
\begin{equation}\label{eq:steom}
	\frac{d^2\varphi}{dx^2}=\frac{dV}{d\varphi}.
\end{equation}
This second order ordinary differential equation can be easily transformed into the first order differential equation
\begin{equation}\label{eq:eqmo_BPS}
    \frac{d\varphi}{dx} = \pm\sqrt{2V}.
\end{equation}

The kinks and antikinks of the model are solutions of Eq.~\eqref{eq:eqmo_BPS} that interpolate between neighboring vacua of the model [i.e., connect adjacent minima of the potential \eqref{eq:potential}]. This means that
\begin{equation}\label{eq:asymptotics}
\varphi(-\infty) \equiv \lim_{x \to -\infty} \varphi (x) = \varphi_1^{\rm (vac)} \quad
\mbox{and} \quad
\varphi(+\infty) \equiv \lim_{x \to +\infty} \varphi (x) = \varphi_2^{\rm (vac)},
\end{equation}
where $\varphi_1^{\rm (vac)}$ and $\varphi_2^{\rm (vac)}$ are two neighboring minima of the potential \eqref{eq:potential}. A static solution having these asymptotics is called a configuration belonging to the topological sector $(\varphi_1^{\rm (vac)},\varphi_2^{\rm (vac)})$. The potential \eqref{eq:potential} has four degenerate minima, $\varphi^{\rm (vac)}=\pm a$ and $\varphi^{\rm (vac)}=\pm b$, hence there are three topological sectors: $(-b,-a)$, $(-a,a)$, and $(a,b)$. As always, the terms ``kink'' and ``antikink'' stand for configurations described by increasing and decreasing functions of coordinate, respectively.

It is important to notice that for non-negative potential \eqref{eq:potential} we can introduce the superpotential $W(\varphi)$ --- a smooth function of $\varphi$ such as
\begin{equation}\label{eq:dwdfi}
V(\varphi) = \frac{1}{2}\left(\frac{dW}{d\varphi}\right)^2.
\end{equation}
Then the energy \eqref{eq:stenergy} of a time-independent configuration can be rewritten as the following:
\begin{equation}\label{eq:static_energy_with_bps}
E_{\rm static}[\varphi] = E_{\rm BPS}^{}[\varphi] + \frac{1}{2}\int_{-\infty}^{\infty}\left(\frac{d\varphi}{dx}\pm\frac{dW}{d\varphi}\right)^2dx,
\end{equation}
with
\begin{equation}\label{eq:static_energy_bps}
E_\mathrm{BPS}^{} = \big|W[\varphi(+\infty)]-W[\varphi(-\infty)]\big|.
\end{equation}
From Eq.~\eqref{eq:static_energy_with_bps} one can see that a static configuration belonging to a given topological sector has the minimal energy if the integrand vanishes, i.e.,
\begin{equation}\label{eq:bps_with_superpotential}
\displaystyle\frac{d\phi}{dx} = \pm\frac{dW}{d\phi}.
\end{equation}
This equation obviously coincides with Eq.~\eqref{eq:eqmo_BPS}. Solutions of Eq.~\eqref{eq:eqmo_BPS}, i.e., kinks and antikinks which are called BPS configurations (or BPS-saturated solutions) \cite{BPS1,BPS2}, have the minimal energy among all possible field configurations in a given topological sector. The energy \eqref{eq:static_energy_bps} is also called kink's (antikink's) {\it mass}.

For the model under consideration with the potential \eqref{eq:potential} we can take the superpotential in the form
\begin{equation}\label{eq:superpotential}
    W(\varphi) = \frac{1}{5}\varphi^5 - \frac{1}{3}\left(a^2+b^2\right)\varphi^3 + a^2b^2\varphi.
\end{equation}
Then the masses of all kinks and antikinks are
\begin{equation}\label{eq:mass_ab}
    M_{(a,b)}^{} = M_{(-b,-a)}^{} = \frac{2(b-a)^3(a^2+3ab+b^2)}{15}
\end{equation}
and
\begin{equation}\label{eq:mass_aa}
    M_{(-a,a)}^{} = \frac{4a^3(5b^2-a^2)}{15}.
\end{equation}
At $a=0$ the mass of the kink in the sector $(-a,a)$ vanishes, while in the sector $(a,b)$ from Eq.~\eqref{eq:mass_ab} we obtain $\displaystyle\frac{2b^5}{15}$, which coincides with \cite[Eq.~(11)]{Belendryasova.CNSNS.2019} taking into account the difference in the definition of the potential \eqref{eq:potential} and the potential \cite[Eq.~(8)]{Belendryasova.CNSNS.2019}. Besides that, the masses \eqref{eq:mass_ab} and \eqref{eq:mass_aa} coincide with the results of  \cite[Sec.~3.1]{Gani.JHEP.2015} at $\lambda=1/\sqrt{2}$. Moreover, the mass of the kink in the sector $(a,b)$ quite naturally vanishes at $a=b$.

Until now, it was believed that kinks of the $\varphi^8$ model can be obtained only in an implicit form, i.e., in the form of the dependence $x=x(\varphi)$ (apparently with the exception of a particular case \cite{Bazeia.AP.2018}). However, as we will demonstrate below, a detailed analysis of the solutions of the static equation of motion shows that, at least for particular values of the ratio $b/a$, kinks can be obtained in the explicit form $\varphi=\varphi(x)$.

\section{Explicit kinks}
\label{sec:kinks}

First, consider \underline{topological sectors $(a,b)$ and $(-b,-a)$}. Substituting the potential \eqref{eq:potential} into the equation of motion \eqref{eq:eqmo_BPS} and integrating with taking into account that $0<a<|\varphi|<b$, we obtain an implicit kink solution:
\begin{equation}
    x = \frac{1}{2\left(b^2-a^2\right)}\ln\left[\left(\frac{\varphi-a}{\varphi+a}\right)^{1/a}\left(\frac{b+\varphi}{b-\varphi}\right)^{1/b}\right].
\end{equation}
For future convenience we transform this equation to the following form:
\begin{equation}
    \left(\frac{\varphi-a}{\varphi+a}\right)^{b/a} \frac{b+\varphi}{b-\varphi} = \exp\left[2b\left(b^2-a^2\right)x\right].
\end{equation}
Then, denoting $b/a=n$ and setting $b=1$, as well as introducing
\begin{equation}
\label{eq:alpha_n}
    \alpha_n(x)=\exp\left[2\left(1-\frac{1}{n^2}\right)x\right],
\end{equation}
we obtain
\begin{equation}
\label{eq:main_algebraic_n}
    \left(\frac{n\:\varphi-1}{n\:\varphi+1}\right)^n \frac{1+\varphi}{1-\varphi} = \alpha_n(x).
\end{equation}
Recall that we consider the case $0<a<b$ and set $b=1$, which can always be ensured by a suitable choice of units for the coordinates $x$, $t$ and for the field $\varphi$. Thus, $n$ can take values in the range $n>1$. The limiting case $n=1$ corresponds to the potential \eqref{eq:potential} with only two minima, which are fourth-order zeros. Note that in this case, the kink connecting the minima will have power-law asymptotics, see \cite[Sec.~II.A]{Christov.PRD.2019}.

For {\it positive integer values of $n$}, the equation \eqref{eq:main_algebraic_n} is an algebraic equation. The problem of solving it reduces to finding the roots of a polynomial of degree $n+1$, with coefficients being dependent on $x$ as a parameter (notice that $\alpha_n(x)>0$). Note that, despite the first paragraph of this section, for even values of $n$ Eq.~\eqref{eq:main_algebraic_n} describes kinks in all topological sectors, while for odd values of $n$ this equation assumes $(\varphi-a)/(\varphi+a)>0$, i.e., $a<|\varphi|<b$. The mass of each kink/antikink in the sectors $(a,b)$ and $(-b,-a)$ [i.e.\ in the sectors $(\frac{1}{n},1)$ and $(-1,-\frac{1}{n})$] as a function of $n$ is
\begin{equation}\label{eq:mass_ab_n}
    M_{(\frac{1}{n},1)}^{} = M_{(-1,-\frac{1}{n})}^{} = \frac{2}{15} \frac{(n-1)^3(n^2+3n+1)}{n^5}.
\end{equation}
In the limit $n\to\infty$ Eq.~\eqref{eq:mass_ab_n} yields $M_{(\frac{1}{n},1)}^{}\to\displaystyle\frac{2}{15}$, see also Eq.~\eqref{eq:mass_ab} and the paragraph below Eq.~\eqref{eq:mass_aa}.

\begin{figure}
    \centering
    \includegraphics[width=0.6
 \textwidth]{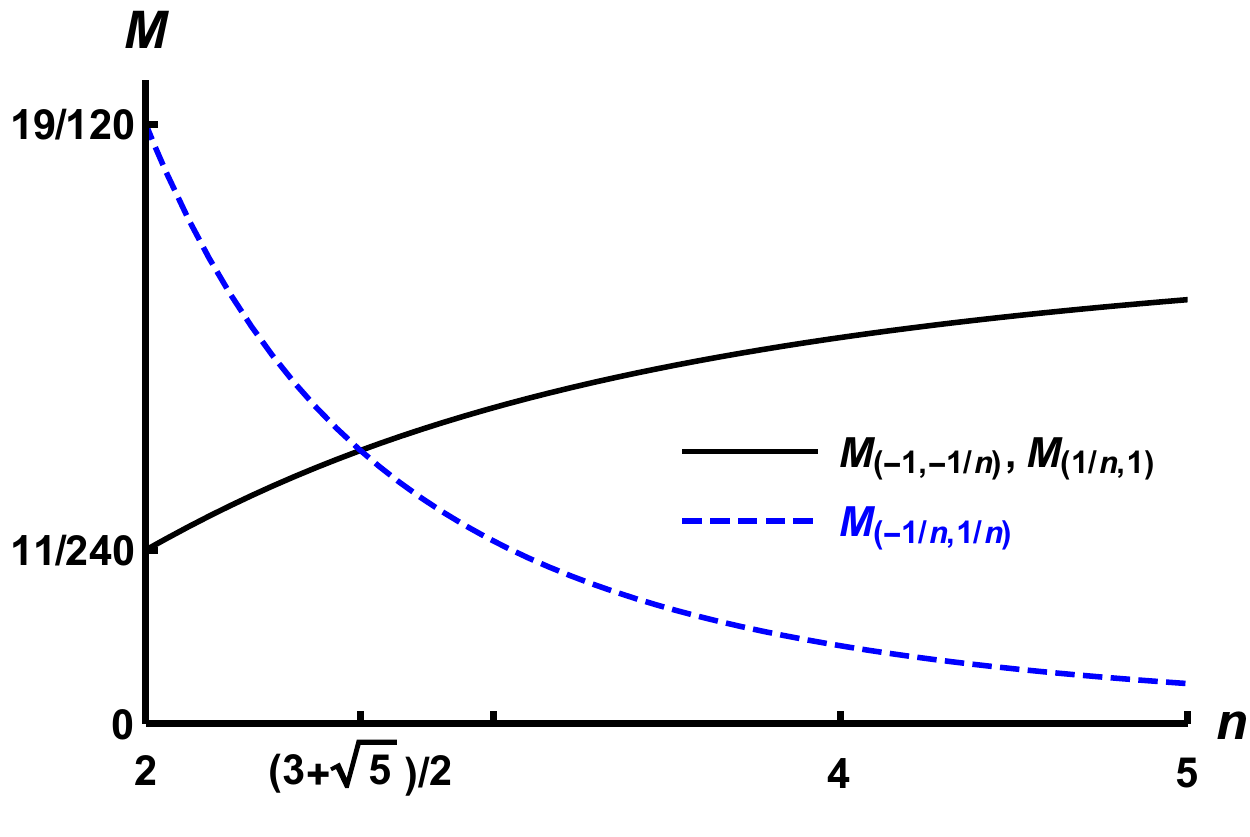}
 \\
    \caption{Masses of kinks as functions of $n=b/a$.}
    \label{fig:Mass}
\end{figure}

In the \underline{topological sector $(-a,a)$} we have $|\varphi|<a$ and therefore we obtain the following algebraic equation
\begin{equation}
\label{eq:main_algebraic_aa_n}
    \left(\frac{1+n\:\varphi}{1-n\:\varphi}\right)^n \frac{1-\varphi}{1+\varphi} = \alpha_n(x),
\end{equation}
and the kink/antikink mass
\begin{equation}\label{eq:mass_aa_n}
    M_{(-\frac{1}{n},\frac{1}{n})}^{} = \frac{4}{15} \frac{5n^2-1}{n^5}.
\end{equation}
In the limit $n\to\infty$ Eq.~\eqref{eq:mass_aa_n} yields obvious result $M_{(-\frac{1}{n},\frac{1}{n})}^{}\to 0$. In Fig.~\ref{fig:Mass}
we show the dependences \eqref{eq:mass_ab_n} and \eqref{eq:mass_aa_n}. It is curious to notice that the two lines intersect at $n=\displaystyle\frac{3+\sqrt{5}}{2}$, which is the square of the golden ratio. The intersection of the curves in Fig.~\ref{fig:Mass} means, that the masses of the kinks in all topological sectors become equal. This, in turn, could have some consequences, e.g., in the kink-(anti)kink collisions. For example, equal masses would enable kinks from different topological sectors to easily transform into one another.

Since we are considering the model with three topological sectors, it is obvious that $a\neq b$. Therefore, the minimum value of $n$ is 2, with which we begin our consideration.

\subsection{The case $n=2$}

\begin{figure}
    \centering
    \includegraphics[width=0.6
 \textwidth]{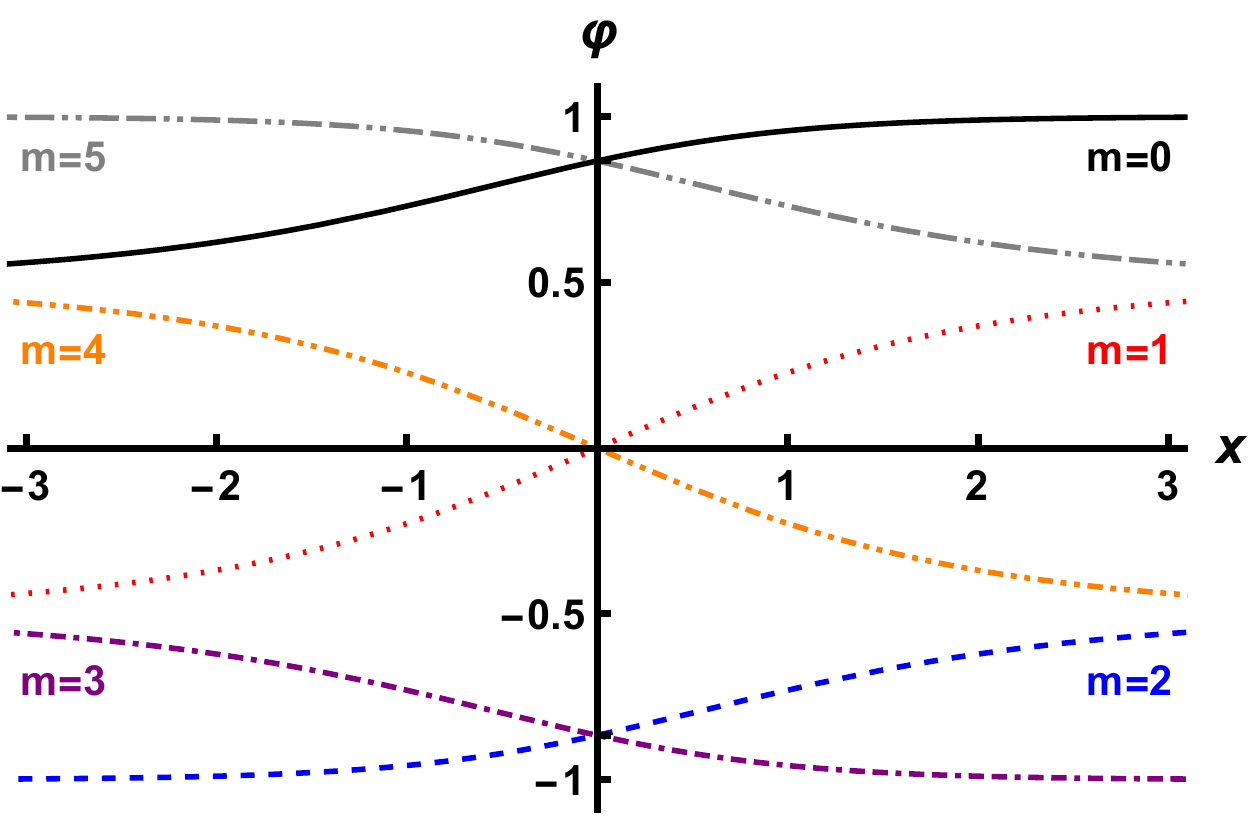}
    \caption{All kinks and antikinks at $n=2$, Eq.~\eqref{eq:kinks_2}, for different values of $m$.} 
    \label{fig:kinksn2}
\end{figure}

At $n=2$ equation \eqref{eq:main_algebraic_n} looks like
\begin{equation}
\label{eq:main_algebraic_2}
4\varphi^3-3\varphi-\beta_2(x) = 0,
\end{equation}
where
\begin{equation}
\label{eq:beta_2}
    \beta_2(x) = \frac{\alpha_2(x)-1}{\alpha_2(x)+1} = \tanh\left(\frac{3}{4}\:x\right).
\end{equation}
To solve equation \eqref{eq:main_algebraic_2} one can substitute $\varphi=\cos\psi$. After some simple algebra, we get the expression for $\varphi(x)$:
\begin{equation}\label{eq:kinks_2}
    \varphi_{\rm K}^{(2)}(x) = \cos\left(\frac{1}{3}\arccos\left[\tanh\left(\frac{3}{4}\:x\right)\right]+\frac{\pi m}{3}\right),
\end{equation}
where $m=0,1,2,3,4,5$ (or takes any other six consecutive integer values). At $m=0$ and $m=5$ Eq.~\eqref{eq:kinks_2} gives kink and antikink in the sector $(\frac{1}{2},1)$; at $m=1$ and $m=4$ Eq.~\eqref{eq:kinks_2} gives kink and antikink in the sector $(-\frac{1}{2},\frac{1}{2})$; at $m=2$ and $m=3$ Eq.~\eqref{eq:kinks_2} gives kink and antikink in the sector $(-1,-\frac{1}{2})$. All these kinks are shown in Fig.~\ref{fig:kinksn2}.

Using explicit formula \eqref{eq:kinks_2} for the kink solutions, we can calculate masses of these kinks/antikinks. To do this, substitute kinks \eqref{eq:kinks_2} into Eq.~\eqref{eq:stenergy} and taking into account Eq.~\eqref{eq:eqmo_BPS} we get
\begin{equation}\label{eq:mass_static_kink}
    M_{\rm K}^{} = E_{\rm static}[\varphi_{\rm K}^{}(x)] = \int_{-\infty}^{\infty}\left( \frac{d\varphi_{\rm K}^{}}{d x} \right)^2 dx = 2\int_{-\infty}^{\infty}V(\varphi_{\rm K}^{}(x))\:dx,
\end{equation}
which yields for $\varphi_{\rm K}^{}(x)=\varphi_{\rm K}^{(2)}(x)$:
\begin{equation}
    M_{(\frac{1}{2},1)}^{} = M_{(-1,-\frac{1}{2})}^{} = \frac{11}{240} \quad \mbox{and} \quad M_{(-\frac{1}{2},\frac{1}{2})}^{} = \frac{19}{120}.
\end{equation}
It is easy to see, that this is exactly the same that could be obtained from Eqs.~\eqref{eq:mass_ab_n}, \eqref{eq:mass_aa_n} for $n=2$.

\subsection{The case $n=3$}

\begin{figure}
    \centering
    \includegraphics[width=0.6
 \textwidth]{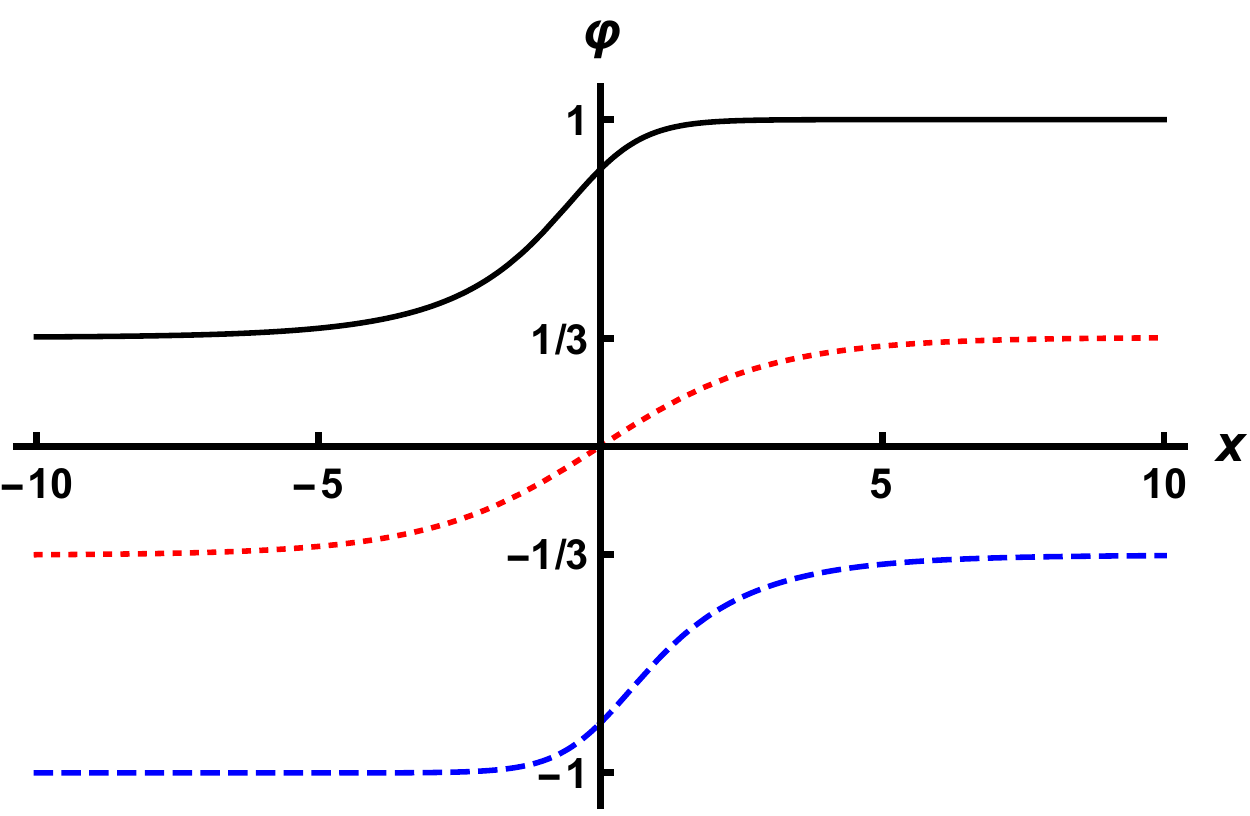}
    \caption{All kinks at $n=3$: Eq.~\eqref{eq:kinks_3_ab} in the sector $(\frac{1}{3},1)$ --- black solid line, Eq.~\eqref{eq:kinks_3_ba} in the sector $(-1,-\frac{1}{3})$ --- blue dashed line, and Eq.~\eqref{eq:kink_3_aa} in the sector $(-\frac{1}{3},\frac{1}{3})$ --- red dotted line.} 
    \label{fig:kinksn3}
\end{figure}

\underline{Topological sectors $(\frac{1}{3},1)$ and $(-1,-\frac{1}{3})$.} At $n=3$ equation \eqref{eq:main_algebraic_n} looks like
\begin{equation}\label{eq:main_algebraic_3}
    27\varphi^4 - 18\varphi^2 - 8\beta_3(x)\varphi - 1 = 0,
\end{equation}
where
\begin{equation}\label{eq:beta_3}
    \beta_3(x) = \frac{\alpha_3(x)-1}{\alpha_3(x)+1} = \tanh\left(\frac{8}{9}\:x\right).
\end{equation}
The equation \eqref{eq:main_algebraic_3} can be solved analytically, i.e., the dependence $\varphi(x)$ can be obtained for all real values of $x$. After some cumbersome but not too complicated algebra, we get the following explicit expressions for kinks:\\
(i) in the topological sector $(\frac{1}{3},1)$
\begin{equation}\label{eq:kinks_3_ab}
    \varphi_{\rm K}^{(3)}(x) = \begin{cases}
        \displaystyle\frac{1}{3} \left(-\sqrt{1-\text{sech}^\frac{2}{3}\left(\frac{8}{9}\:x\right)}+\sqrt{2+\text{sech}^{\frac{2}{3}}\left(\frac{8}{9}\:x\right)-\frac{2\tanh \left(\frac{8}{9}\:x\right)}{\sqrt{1-\text{sech}^\frac{2}{3}\left(\frac{8}{9}\:x\right)}}}\right), \quad x<0,\\
        \displaystyle\frac{1}{3} \left(\sqrt{1-\text{sech}^\frac{2}{3}\left(\frac{8}{9}\:x\right)}+\sqrt{2+\text{sech}^{\frac{2}{3}}\left(\frac{8}{9}\:x\right)+\frac{2\tanh \left(\frac{8}{9}\:x\right)}{\sqrt{1-\text{sech}^\frac{2}{3}\left(\frac{8}{9}\:x\right)}}}\right), \quad x>0,
    \end{cases}
\end{equation}
and (ii) in the topological sector $(-1,-\frac{1}{3})$
\begin{equation}\label{eq:kinks_3_ba}
    \varphi_{\rm K}^{(3)}(x) = \begin{cases}
       \displaystyle\frac{1}{3} \left(-\sqrt{1-\text{sech}^\frac{2}{3}\left(\frac{8}{9}\:x\right)}-\sqrt{2+\text{sech}^{\frac{2}{3}}\left(\frac{8}{9}\:x\right)-\frac{2\tanh \left(\frac{8}{9}\:x\right)}{\sqrt{1-\text{sech}^\frac{2}{3}\left(\frac{8}{9}\:x\right)}}}\right), \quad x<0,\\
       \displaystyle\frac{1}{3} \left(\sqrt{1-\text{sech}^\frac{2}{3}\left(\frac{8}{9}\:x\right)}-\sqrt{2+\text{sech}^{\frac{2}{3}}\left(\frac{8}{9}\:x\right)+\frac{2\tanh \left(\frac{8}{9}\:x\right)}{\sqrt{1-\text{sech}^\frac{2}{3}\left(\frac{8}{9}\:x\right)}}}\right), \quad x>0.
    \end{cases}
\end{equation}
The functions \eqref{eq:kinks_3_ab}, \eqref{eq:kinks_3_ba} are not defined at the point $x=0$, nevertheless, they have pairwise equal one-sided limits, namely
\begin{equation}
    \lim_{x \to -0} \varphi_{\rm K}^{(3)}(x) = \lim_{x \to +0} \varphi_{\rm K}^{(3)}(x) = \frac{\sqrt{3+2\sqrt{3}}}{3} \quad
\mbox{in the sector $(\frac{1}{3},1)$}
\end{equation}
and
\begin{equation}
    \lim_{x \to -0} \varphi_{\rm K}^{(3)}(x) = \lim_{x \to +0} \varphi_{\rm K}^{(3)}(x) = -\frac{\sqrt{3+2\sqrt{3}}}{3} \quad
\mbox{in the sector $(-1,-\frac{1}{3})$}.
\end{equation}
Both kinks \eqref{eq:kinks_3_ab} and \eqref{eq:kinks_3_ba} are plotted in Fig.~\ref{fig:kinksn3}.

\underline{Topological sector $(-\frac{1}{3},\frac{1}{3})$.} In order to obtain kink solution in this sector, we have to solve Eq.~\eqref{eq:main_algebraic_aa_n} with $n=3$, i.e.,
\begin{equation}\label{eq:main_algebraic_3_aa}
    27\varphi^4 - 18\varphi^2 + \frac{8}{\beta_3(x)}\varphi - 1 = 0,
\end{equation}
where $\beta_3(x)$ is defined by Eq.~\eqref{eq:beta_3} above. Equation \eqref{eq:main_algebraic_3_aa} can be solved quite similarly to Eq.~\eqref{eq:main_algebraic_3}. As a result we get the following explicit formula for kink in the topological sector $(-\frac{1}{3},\frac{1}{3})$:
\begin{equation}\label{eq:kink_3_aa}
    \varphi_{\rm K}^{(3)}(x) = \begin{cases}
       \displaystyle\frac{1}{3} \left(\sqrt{1+\text{csch}^\frac{2}{3}\left(\frac{8}{9}\:x\right)}-\sqrt{2-\text{csch}^{\frac{2}{3}}\left(\frac{8}{9}\:x\right)-\frac{2\coth \left(\frac{8}{9}\:x\right)}{\sqrt{1+\text{csch}^\frac{2}{3}\left(\frac{8}{9}\:x\right)}}}\right), \quad x<0,\\       \displaystyle\frac{1}{3} \left(-\sqrt{1+\text{csch}^\frac{2}{3}\left(\frac{8}{9}\:x\right)}+\sqrt{2-\text{csch}^{\frac{2}{3}}\left(\frac{8}{9}\:x\right)+\frac{2\coth \left(\frac{8}{9}\:x\right)}{\sqrt{1+\text{csch}^\frac{2}{3}\left(\frac{8}{9}\:x\right)}}}\right), \quad x>0,
    \end{cases}
\end{equation}
The kink \eqref{eq:kink_3_aa} is shown in Fig.~\ref{fig:kinksn3}.
Note that, as in the previous case of the topological sectors $(\frac{1}{3},1)$ and $(-1,-\frac{1}{3})$, both functions in Eq.~\eqref{eq:kink_3_aa} are also not defined at the point $x=0$, however, they have equal one-sided limits:
\begin{equation}
    \lim_{x \to -0} \varphi_{\rm K}^{(3)}(x) = \lim_{x \to +0} \varphi_{\rm K}^{(3)}(x) = 0 \quad
\mbox{for the kink in the sector $(-\frac{1}{3},\frac{1}{3})$}.
\end{equation}

Using the obtained explicit expressions for kinks in all topological sectors for $n=3$, we can calculate the kink's masses. Substituting Eqs.~\eqref{eq:kinks_3_ab}, \eqref{eq:kinks_3_ba} and \eqref{eq:kink_3_aa} into Eq.~\eqref{eq:mass_static_kink}, we get
\begin{equation}
    M_{(\frac{1}{3},1)}^{} = M_{(-1,-\frac{1}{3})}^{} = \frac{304}{3645} \quad \mbox{and} \quad M_{(-\frac{1}{3},\frac{1}{3})}^{} = \frac{176}{3645}.
\end{equation}
Surely, the same values could also be obtained from Eqs.~\eqref{eq:mass_ab_n} and \eqref{eq:mass_aa_n} at $n=3$.

As for $n>3$, some features of the polynomial equations obtained in the cases $n=2$ and $n=3$ give reason to hope that, for some $n>3$, explicit solutions can also be found. On the other hand, if we compare solutions \eqref{eq:kinks_2} for $n=2$ and \eqref{eq:kinks_3_ab}, \eqref{eq:kinks_3_ba}, \eqref{eq:kink_3_aa} for $n=3$, then we can assume that for larger $n$ explicit solutions (if any) will be extremely cumbersome.

\subsection{Asymptotics of kinks}

The asymptotics of kinks for any positive integer $n>2$ can be obtained from Eqs.~\eqref{eq:main_algebraic_n} and \eqref{eq:main_algebraic_aa_n}. We consider separately the topological sectors $(\frac{1}{n},1)$ (asymmetric kink) and $(-\frac{1}{n},\frac{1}{n})$ (symmetric kink).

\underline{Topological sector $(\frac{1}{n},1)$.} At $x\to-\infty$ we have $\varphi_{\rm K}^{(n)}(x)\to\displaystyle\frac{1}{n}+0$, i.e.,
\begin{equation}\label{eq:delta_phi_1}
\varphi_{\rm K}^{(n)}(x) = \frac{1}{n} + \delta\varphi,
\end{equation}
where $\delta\varphi>0$, $|\delta\varphi|\ll 1$. Substituting Eq.~\eqref{eq:delta_phi_1} into Eq.~\eqref{eq:main_algebraic_n} and linearizing with respect to $\delta\varphi$, we obtain:
\begin{equation}\label{eq:delta_phi_1_1}
\delta\varphi(x) \approx \frac{2}{n}\left(\frac{n-1}{n+1}\right)^{\frac{1}{n}}\exp{\left[\frac{2}{n}\left(1-\frac{1}{n^{2}}\right)x\right]}.
\end{equation}
At $x\to+\infty$ we have $\varphi_{\rm K}^{(n)}(x)\to 1-0$, i.e.,
\begin{equation}\label{eq:delta_phi_2}
\varphi_{\rm K}^{(n)}(x) = 1 - \delta\varphi,
\end{equation}
where $\delta\varphi>0$, $|\delta\varphi|\ll 1$. Substituting Eq.~\eqref{eq:delta_phi_2} into Eq.~\eqref{eq:main_algebraic_n} and linearizing with respect to $\delta\varphi$, we obtain:
\begin{equation}
\delta\varphi(x) \approx 2\left(\frac{n-1}{n+1}\right)^{n}\exp{\left[-2\left(1-\frac{1}{n^{2}}\right)x\right]}.
\end{equation}
Thus, we obtain the asymptotics of the asymmetric kink in the topological sector $(\frac{1}{n},1)$:
\begin{equation}
\varphi_{\rm K}^{(n)}(x) \approx \begin{cases}
    \displaystyle\frac{1}{n} + \frac{2}{n}\left(\frac{n-1}{n+1}\right)^{\frac{1}{n}}\exp{\left[\frac{2}{n}\left(1-\frac{1}{n^{2}}\right)x\right]} \quad \mbox{at} \quad x \to -\infty,\\
    \displaystyle 1 - 2\left(\frac{n-1}{n+1}\right)^{n}\exp{\left[-2\left(1-\frac{1}{n^{2}}\right)x\right]} \quad \mbox{at} \quad x \to +\infty.
\end{cases}
\end{equation}
In particular, at $n=2$ the above formulas give us asymptotics of the kink \eqref{eq:kinks_2} at $m=0$:
\begin{equation}
\varphi_{\rm K}^{(2)}(x) \approx \begin{cases}
    \displaystyle\frac{1}{2} + \frac{1}{\sqrt{3}}\exp\left(\frac{3}{4}\:x\right) \quad \mbox{at} \quad x \to -\infty,\\
    1 - \displaystyle\frac{2}{9}\exp\left(-\frac{3}{2}\:x\right) \quad \mbox{at} \quad x \to +\infty.
\end{cases}
\end{equation}
At $n=3$ we obtain asymptotics of the kink \eqref{eq:kinks_3_ab}:
\begin{equation}
\varphi_{\rm K}^{(3)}(x) \approx \begin{cases}
    \displaystyle\frac{1}{3} +  \frac{2^{2/3}}{3}\exp\left(\frac{16}{27}\:x\right) \quad \mbox{at} \quad x \to -\infty,\\
    1 - \displaystyle\frac{1}{4}\exp\left(-\frac{16}{9}\:x\right) \quad \mbox{at} \quad x \to +\infty.
    \end{cases}
\end{equation}

In the limit $n\to\infty$ (i.e., for $a\to 0$) at $x\to-\infty$ it can be seen that for any finite $x$ the argument of the exponent tends to zero, which corresponds to the transition from exponential to power-law asymptotic behavior of the kink in the topological sector $(0,1)$, cf.~\cite[Eq.~(22)]{Christov.PRD.2019}. At $x\to+\infty$ we obtain:
\begin{equation}
\varphi_{\rm K}^{(\infty)}(x) \approx \, 1 -  \frac{2}{e^2}\,e^{-2x},
\end{equation}
which can be compared with \cite[Eq.~(23)]{Christov.PRD.2019}, taking into account difference in the choice of the model potentials.

\underline{Topological sector $(-\frac{1}{n},\frac{1}{n})$.} At $x\to-\infty$ we have $\varphi_{\rm K}^{(n)}(x)\to-\displaystyle\frac{1}{n}+0$, i.e.,
\begin{equation}\label{eq:delta_phi_3}
\varphi_{\rm K}^{(n)}(x) = -\frac{1}{n} + \delta\varphi,
\end{equation}
where $\delta\varphi>0$, $|\delta\varphi|\ll 1$. Substituting Eq.~\eqref{eq:delta_phi_3} into Eq.~\eqref{eq:main_algebraic_aa_n} and linearizing with respect to $\delta\varphi$, we obtain:
\begin{equation}\label{eq:delta_phi_2_2}
\delta\varphi(x) \approx \frac{2}{n}\left(\frac{n-1}{n+1}\right)^{\frac{1}{n}}\exp{\left[\frac{2}{n}\left(1-\frac{1}{n^{2}}\right)x\right]},
\end{equation}
which, of course, coincides with \eqref{eq:delta_phi_1_1}. Taking into account symmetry of the kink in the sector $(-\frac{1}{n},\frac{1}{n})$, we get the kink's asymptotics:
\begin{equation}
\varphi_{\rm K}^{(n)}(x) \approx \begin{cases}
    -\displaystyle\frac{1}{n} + \frac{2}{n}\left(\frac{n-1}{n+1}\right)^{\frac{1}{n}}\exp{\left[\frac{2}{n}\left(1-\frac{1}{n^{2}}\right)x\right]} \quad \mbox{at} \quad x \to -\infty,\\
    \displaystyle\frac{1}{n} - \frac{2}{n}\left(\frac{n-1}{n+1}\right)^{\frac{1}{n}}\exp{\left[-\frac{2}{n}\left(1-\frac{1}{n^{2}}\right)x\right]} \quad \mbox{at} \quad x \to +\infty.
\end{cases}
\end{equation}
In particular, at $n=2$ from the above formulas we get asymptotics of the kink \eqref{eq:kinks_2} at $m=1$:
\begin{equation}
\varphi_{\rm K}^{(2)}(x) \approx \begin{cases}
    -\displaystyle\frac{1}{2} + \frac{1}{\sqrt{3}}\exp\left(\frac{3}{4}\:x\right) \quad \mbox{at} \quad x \to -\infty,\\
    \displaystyle\frac{1}{2} - \frac{1}{\sqrt{3}}\exp\left(-\frac{3}{4}\:x\right) \quad \mbox{at} \quad x \to +\infty,
\end{cases}
\end{equation}
while at $n=3$ we obtain asymptotics of the kink \eqref{eq:kink_3_aa}:
\begin{equation}
\varphi_{\rm K}^{(3)}(x) \approx \begin{cases}
    -\displaystyle\frac{1}{3} +  \frac{2^{2/3}}{3}\exp\left(\frac{16}{27}\:x\right) \quad \mbox{at} \quad x \to -\infty,\\
    \displaystyle\frac{1}{3} -  \frac{2^{2/3}}{3}\exp\left(-\frac{16}{27}\:x\right) \quad \mbox{at} \quad x \to +\infty.
    \end{cases}
\end{equation}
In the limit $n\to\infty$ (i.e., for $a\to 0$) the topological sector $(-\frac{1}{n},\frac{1}{n})$ vanishes.

\subsection{The case of rational $n$}

In the case of rational $n=\displaystyle\frac{p}{q}$, Eq.~\eqref{eq:kink_3_aa} takes the form:
\begin{equation}\label{eq:kink_3_aa_rational}
\left(\frac{p\varphi-q}{p\varphi+q}\right)^{p}\left(\frac{1+\varphi}{1-\varphi}\right)^{q} = \left(\alpha_{p/q}^{}(x)\right)^q,
\end{equation}
where
\begin{equation}
\alpha_{p/q}^{}(x) = \exp{\left[2\left(1-\frac{q^2}{p^2}\right)x \right]},\quad \mbox{hence} \quad \left(\alpha_{p/q}^{}(x)\right)^q = \exp\left[2q\left(1-\frac{q^2}{p^2}\right)x \right].
\end{equation}
Equation \eqref{eq:kink_3_aa_rational} represents polynomial equation of degree $p+q$:
\begin{equation}
(p\varphi-q)^p(1+\varphi)^q = \left(\alpha_{p/q}^{}(x)\right)^q \left(p\varphi+q\right)^p\left(1-\varphi\right)^q.
\end{equation}
For example, at $n=3/2$ ($p=3$, $q=2$) we have
\begin{equation}
(3\varphi-2)^3(1+\varphi)^2 = \left(\alpha_{3/2}^{}(x)\right)^2 \left(3\varphi+2\right)^3\left(1-\varphi\right)^2,
\end{equation}
which is a fifth degree equation for $\varphi$. Solving of such equations is beyond the scope of this paper and could be the subject of a separate study.

\section{Kink's excitation spectra}
\label{sec:properties}

Having at hand explicit formulas for kinks, now we can study the kinks' excitation spectra. The problem is formulated as follows (see, e.g., \cite[Sec.~2]{Gani.JHEP.2015}, \cite[Sec.~4]{Belendryasova.CNSNS.2019}). We add a small perturbation $\delta\varphi(x,t)$ to the static kink $\varphi_{\rm K}^{}(x)$, the excitation spectrum of which we are looking for, i.e.,
\begin{equation}
\varphi(x,t) = \varphi_{\rm K}^{}(x) + \delta\varphi(x,t), \quad ||\delta\varphi|| \ll ||\varphi_{\rm K}^{}||.
\end{equation}
Substituting this $\varphi(x,t)$ into the equation of motion \eqref{eq:eom}, we obtain in a linear approximation in $\delta\varphi$:
\begin{equation}\label{eq:delta_phi}
\frac{\partial^2\delta\varphi}{\partial t^2} - \frac{\partial^2\delta\varphi}{\partial x^2} + \left.\frac{d^2 V}{d\varphi^2}\right|_{\varphi_{\rm K}^{}(x)}\cdot\delta\varphi = 0.
\end{equation}
We can separate the variables $x$ and $t$ or, in other words, we can look for a solution of this equation in the form
\begin{equation}
\delta\varphi(x,t) = \psi(x)\cos\:\omega t.
\end{equation}
Then Eq.~\eqref{eq:delta_phi} yields the following eigenvalue problem:
\begin{equation}\label{eq:stat_Schr}
\hat{H}\psi(x) = \omega^2\psi(x),
\end{equation}
which is similar to the one-dimensional stationary Schr\"odinger equation with the Hamiltonian
\begin{equation}\label{eq:Schr_Ham}
\hat{H} = -\frac{d^2}{dx^2} + U(x).
\end{equation}
The function $U(x)$ is the stability potential which can be viewed as ``quantum-mechanical'' potential. The ``energy levels'' of the discrete spectrum in the potential well $U(x)$ are nothing else than eigenvalues $\omega_i^2$, which are our ultimate goal. It is easy to find that
\begin{equation}\label{eq:Schr_pot}
U(x) = \left.\frac{d^2 V}{d\varphi^2}\right|_{\varphi_{\rm K}^{}(x)}.
\end{equation}
Notice that it can also be easily shown that there is always a zero level in the kink's excitation spectrum, see, e.g., \cite[Eqs.~(25), (26)]{Belendryasova.CNSNS.2019}. Moreover, all eigenvalues of the operator $\hat{H}$ are non-negative \cite[Eqs.~(2.20), (2.21)]{Bazeia.EPJC.2018}, \cite[Sec.~2]{Bazeia.AP.2018}.

For a numerical search of eigenvalues $\omega_i^{}$ of the discrete spectrum we used a modification of the ``shooting method'' (or ``matching method'') \cite[Sec.~9.4]{Izaac.book.2018}, see also \cite[Sec.~IV]{Gani.PRE.1999}, \cite[Sec.~3.1.1]{Gani.JHEP.2015}, \cite[Sec.~4]{Belendryasova.CNSNS.2019}. To be brief, the essence of the method is as follows. The ordinary differential equation \eqref{eq:stat_Schr} at a particular value of $\omega$ is solved numerically separately at $x<0$ and $x>0$ starting from the left and the right infinity, respectively. Then the two obtained solutions $\psi_{\rm L}^{}(x)$ and $\psi_{\rm R}^{}(x)$ are matched in some point $x=x_{\rm match}^{}$ near the origin. If the selected value of $\omega$ is an eigenvalue of the Hamiltonian \eqref{eq:Schr_Ham}, the ``left'' and the ``right'' solutions would be parts of the same eigenfunction of $\hat{H}$. This entails zeroing out the Wronskian of the functions $\psi_{\rm L}^{}(x)$ and $\psi_{\rm R}^{}(x)$ at $x=x_{\rm match}^{}$. To find out the functions $\psi_{\rm L}^{}(x)$ and $\psi_{\rm R}^{}(x)$, we solved the ordinary differential equation \eqref{eq:stat_Schr} numerically using the classic fourth-order Runge-Kutta method with the step $h=10^{-5}$.

Substituting explicit formulas for kinks at $n=2$ and $n=3$ into Eq.~\eqref{eq:Schr_pot}, we can get the ``quantum-mechanical'' potentials for each kink. The final formulas are too bulky, and we do not give them here, however obtaining them either manually or using computer algebra system does not present fundamental difficulties. For convenience, consider the two cases $n=2$ and $n=3$ separately and in more detail.

\begin{figure}
    \centering
    \includegraphics[width=0.6
 \textwidth]{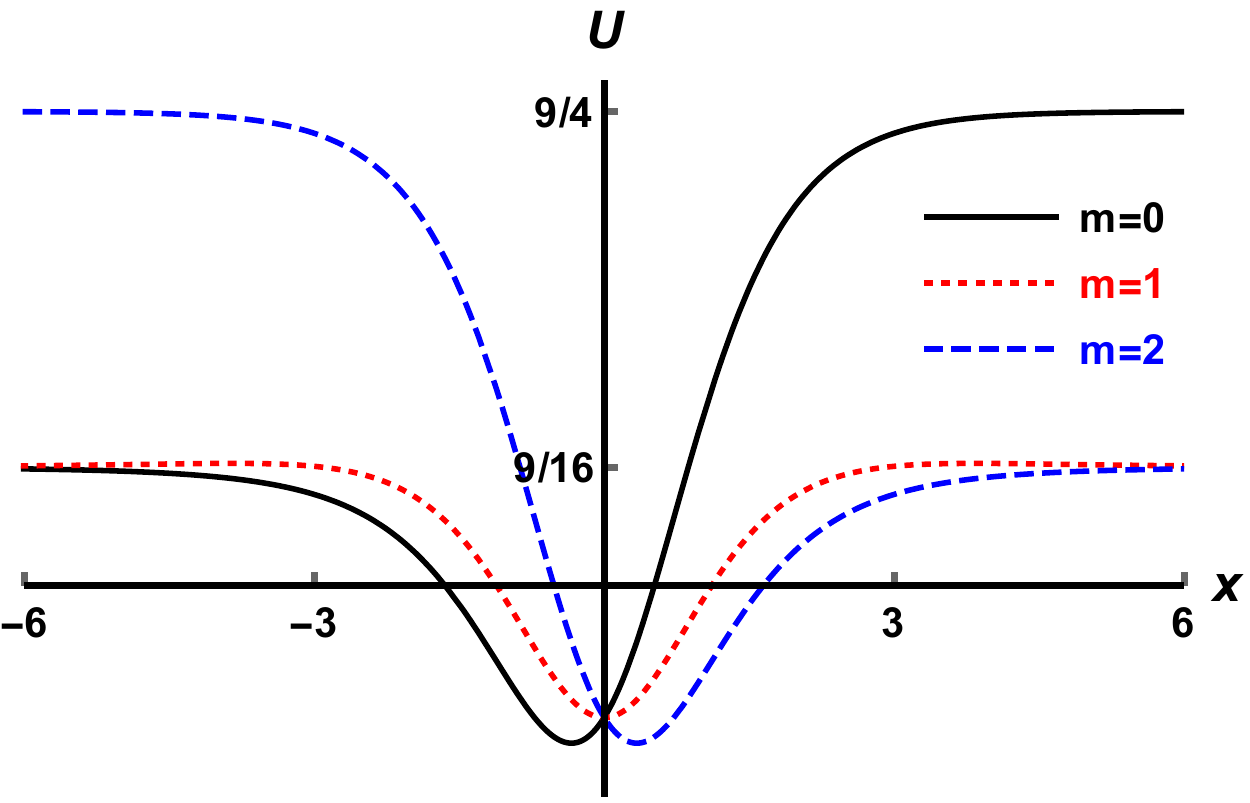}
    \caption{The ``quantum-mechanical'' (stability) potential $U(x)$ for kinks in the case $n=2$ (see Eq.~\eqref{eq:kinks_2} and the text below).} 
    \label{fig:QMPn2}
\end{figure}

\underline{The case $n=2$.} The potentials $U(x)$ for all three kinks are shown in Fig.~\ref{fig:QMPn2}.
For the kinks in the topological sectors $(\frac{1}{2},1)$ and $(-1,-\frac{1}{2})$ the potentials are obviously mirror symmetric, i.e., $U_{(-1,-\frac{1}{2})}^{}(x)=U_{(\frac{1}{2},1)}^{}(-x)$. Therefore, we focus on the kink $(\frac{1}{2},1)$ only. The corresponding potential has asymptotics
\begin{equation}
    U_{(\frac{1}{2},1)}^{}(-\infty) = \lim_{x \to -\infty} U_{(\frac{1}{2},1)}^{}(x) = \left.\frac{d^2V}{d\varphi^2}\right|_{\varphi=\frac{1}{2}} = \frac{9}{16}
\end{equation}
and
\begin{equation}
    U_{(\frac{1}{2},1)}^{}(+\infty) = \lim_{x \to +\infty} U_{(\frac{1}{2},1)}^{}(x) = \left.\frac{d^2V}{d\varphi^2}\right|_{\varphi=1} = \frac{9}{4}.
\end{equation}
Besides, the minimal value $U_{\rm min}^{}\approx -0.74651$ is reached at
%
%
\begin{equation}
x_{\rm min}^{}=\frac{4}{3}\:\mbox{artanh}\left(
\frac{-17+\sqrt{163}}{28}\sqrt{\frac{25+\sqrt{163}}{14}}\right)
\approx -0.33813.
\end{equation}
The discrete spectrum of the operator $\hat{H}$ is thus localized in the range $0\le\omega^2\le 0.5625$. We performed the numerical search for discrete levels in the potential well $U_{(\frac{1}{2},1)}^{}(x)$ and found only the zero mode $\omega_0^2\approx 2\cdot 10^{-13}$.

For the kink in the topological sector $(-\frac{1}{2},\frac{1}{2})$ the ``quantum-mechanical'' potential $U_{(-\frac{1}{2},\frac{1}{2})}^{}(x)$ is symmetric, $U_{(-\frac{1}{2},\frac{1}{2})}^{}(-x)=U_{(-\frac{1}{2},\frac{1}{2})}^{}(x)$, it has asymptotics
\begin{equation}
    U_{(-\frac{1}{2},\frac{1}{2})}^{}(-\infty) = U_{(-\frac{1}{2},\frac{1}{2})}^{}(+\infty) = \lim_{x \to \pm\infty} U_{(-\frac{1}{2},\frac{1}{2})}^{}(x) = \left.\frac{d^2V}{d\varphi^2}\right|_{\varphi=\pm\frac{1}{2}} = \frac{9}{16}
\end{equation}
and the minimal value $U_{\rm min}^{}=-\displaystyle\frac{5}{8}$ at $x=0$. The discrete spectrum of the operator $\hat{H}$ for this kink is localized in the range $0\le\omega^2\le 0.5625$. Numerical search for discrete levels in the potential well $U_{(-\frac{1}{2},\frac{1}{2})}^{}(x)$ gives only the zero mode $\omega_0^2\approx -2\cdot 10^{-14}$.

\begin{figure}
    \centering
    \includegraphics[width=0.6
 \textwidth]{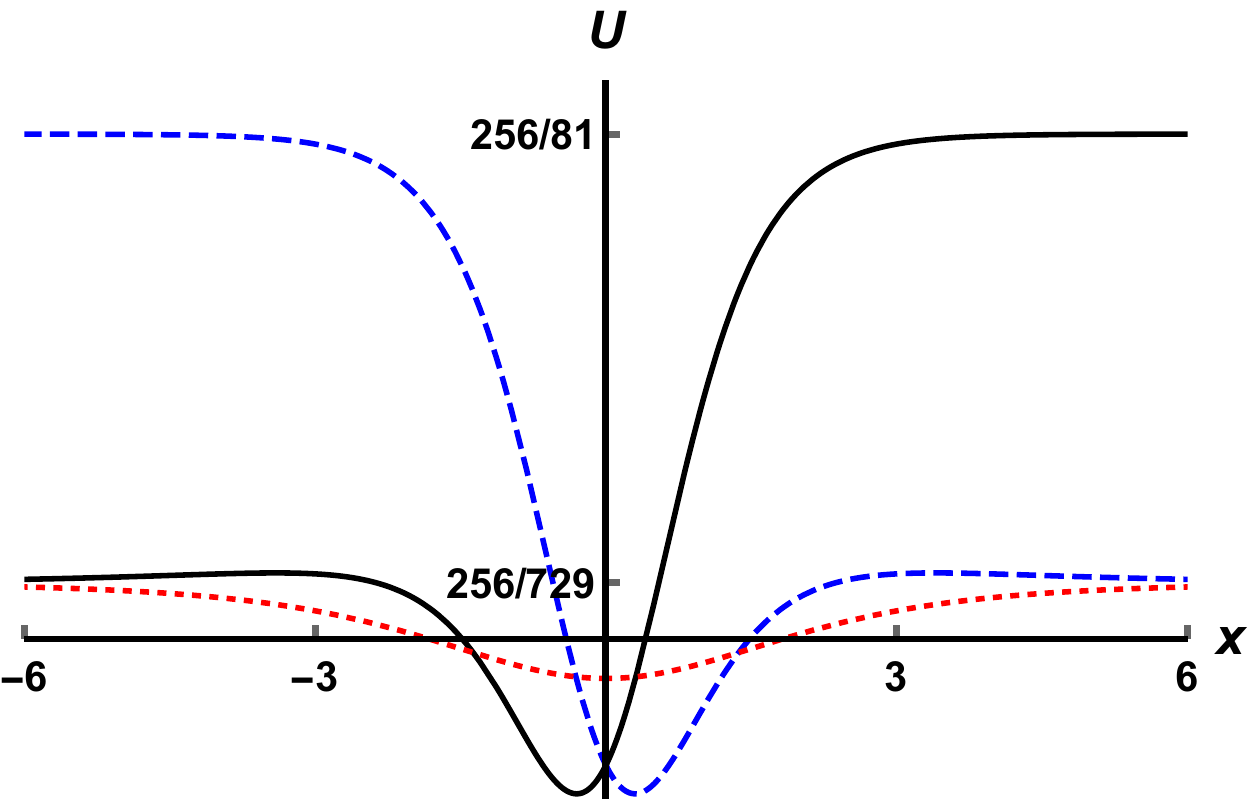}
    \caption{The ``quantum-mechanical'' (stability) potential $U(x)$ for kinks in the case $n=3$.} 
    \label{fig:QMPn3}
\end{figure}

\underline{The case $n=3$.} The potentials $U(x)$ for all three kinks are shown in Fig.~\ref{fig:QMPn3}.
As in the previous case, for the kinks in the topological sectors $(\frac{1}{3},1)$ and $(-1,-\frac{1}{3})$ the potentials are mirror symmetric, $U_{(-1,-\frac{1}{3})}^{}(x)=U_{(\frac{1}{3},1)}^{}(-x)$, so we focus on the kink $(\frac{1}{3},1)$ only. The asymptotics of the corresponding potential are
\begin{equation}
    U_{(\frac{1}{3},1)}^{}(-\infty) = \lim_{x \to -\infty} U_{(\frac{1}{3},1)}^{}(x) = \left.\frac{d^2V}{d\varphi^2}\right|_{\varphi=\frac{1}{3}} = \frac{256}{729}
\end{equation}
and
\begin{equation}
    U_{(\frac{1}{3},1)}^{}(+\infty) = \lim_{x \to +\infty} U_{(\frac{1}{3},1)}^{}(x) = \left.\frac{d^2V}{d\varphi^2}\right|_{\varphi=1} = \frac{256}{81}.
\end{equation}
The minimal value $U_{\rm min}^{}=U_{(\frac{1}{3},1)}^{}(x_{\rm min}^{})\approx -0.969014$ at $x_{\rm min}^{}\approx 0.300959$, and the discrete spectrum of the operator $\hat{H}$ is localized in the range $0\le\omega^2\le 0.351166$. As a result of the numerical search for discrete levels in the potential well $U_{(\frac{1}{3},1)}^{}(x)$ we obtained only the zero mode with frequency $\omega_0^2\approx 5\cdot 10^{-11}$.

The kink in the topological sector $(-\frac{1}{3},\frac{1}{3})$ has the symmetric ``quantum-mechanical'' potential, $U_{(-\frac{1}{3},\frac{1}{3})}^{}(-x)=U_{(-\frac{1}{3},\frac{1}{3})}^{}(x)$, with asymptotics
\begin{equation}
    U_{(-\frac{1}{3},\frac{1}{3})}^{}(-\infty) = U_{(-\frac{1}{3},\frac{1}{3})}^{}(+\infty) = \lim_{x \to \pm\infty} U_{(-\frac{1}{3},\frac{1}{3})}^{}(x) = \left.\frac{d^2V}{d\varphi^2}\right|_{\varphi=\pm\frac{1}{3}} = \frac{256}{729}
\end{equation}
and the minimal value $U_{\rm min}^{}=U_{(-\frac{1}{3},\frac{1}{3})}^{}(0)=-0.246914$. The discrete spectrum of the operator $\hat{H}$ for this kink is localized in the range $0\le\omega^2\le 0.351166$. The numerical search for discrete levels in the potential well $U_{(-\frac{1}{3},\frac{1}{3})}^{}(x)$ has shown only the zero mode with $\omega_0^2\approx 3\cdot 10^{-13}$.

\begin{figure}[h!]
\begin{center}
  \centering
    \subfigure[$n=2$, topological sector ($-\frac{1}{2}, \frac{1}{2}$)]
{\includegraphics[width=0.48\textwidth]{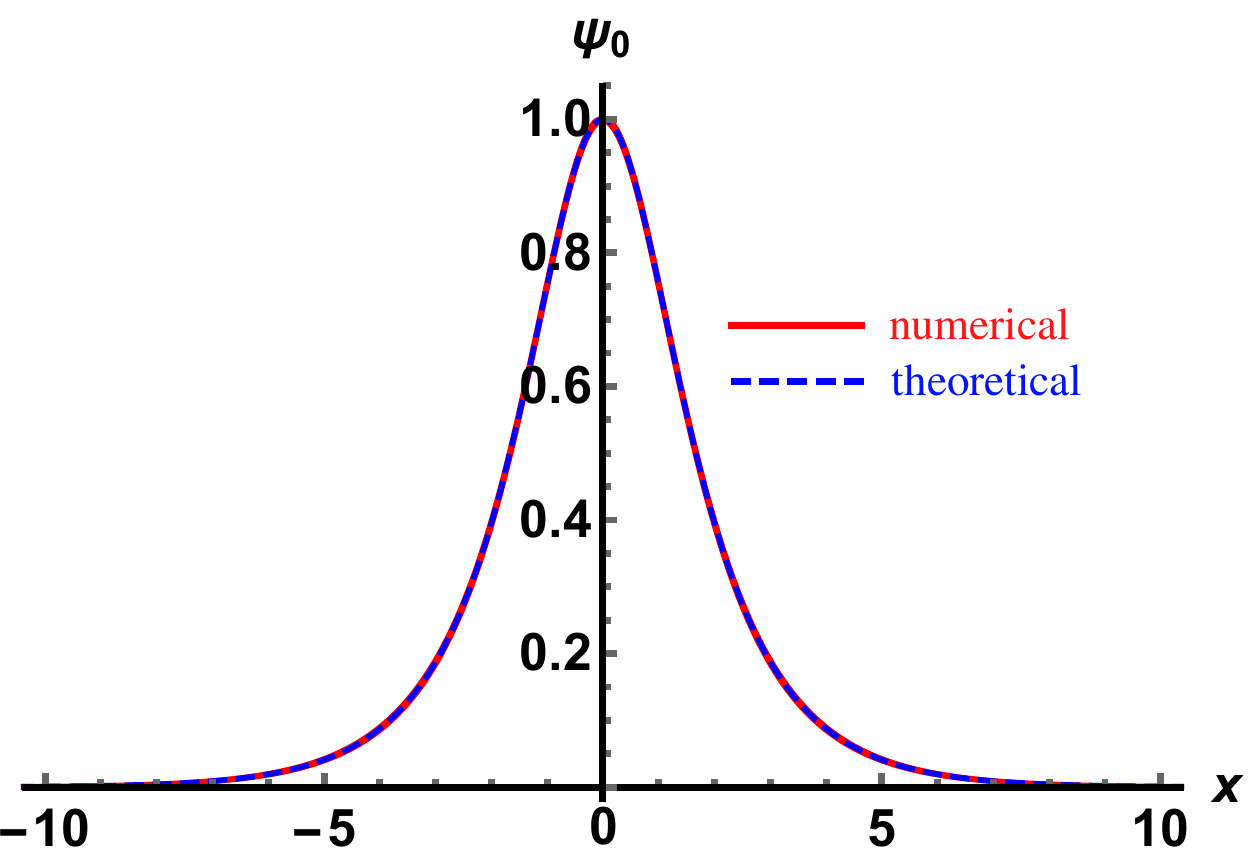}\label{fig:Sn2WaveFunction}}
    \subfigure[$n=2$, topological sector ($\frac{1}{2}, 1$)]
{\includegraphics[width=0.48\textwidth]{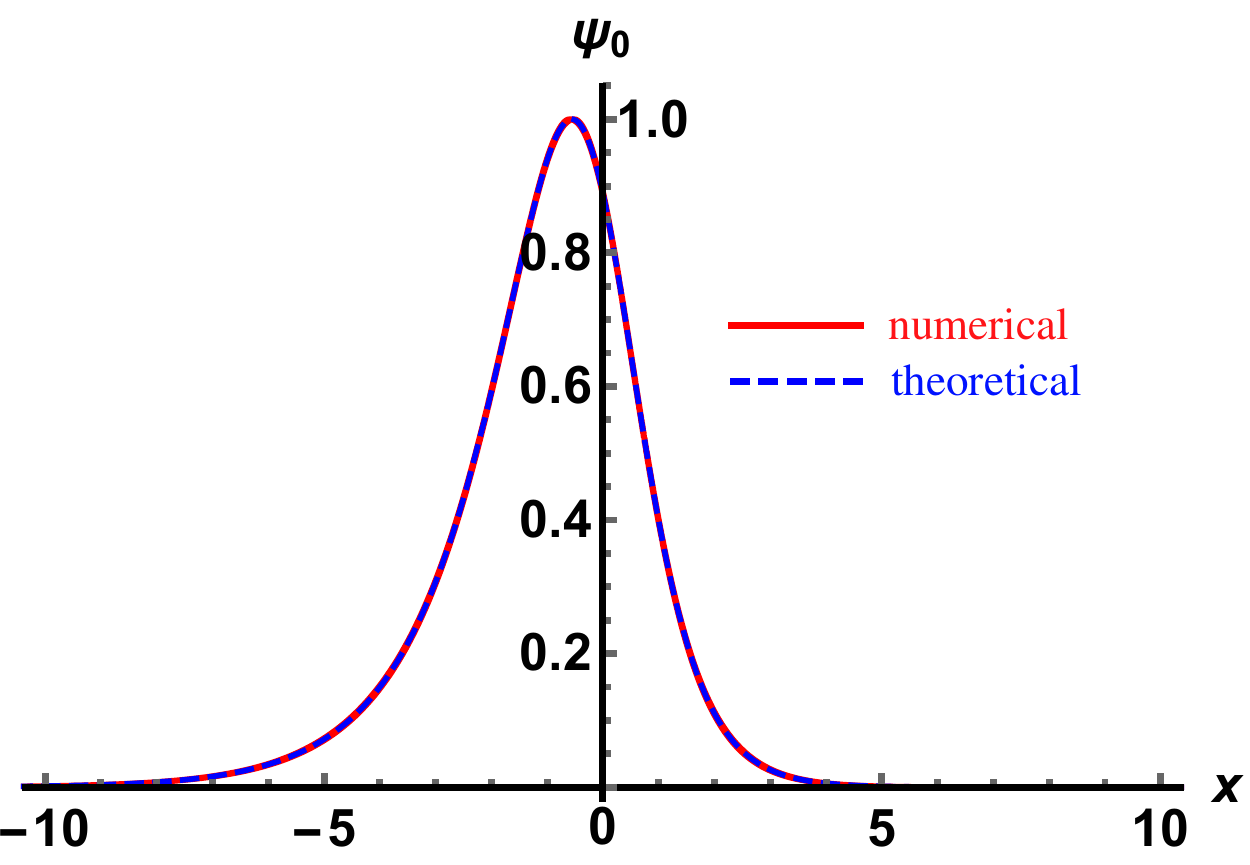}\label{fig:ASn2WaveFunction}}
\\
    \subfigure[$n=3$, topological sector ($-\frac{1}{3}, \frac{1}{3}$)]
{\includegraphics[width=0.48\textwidth]{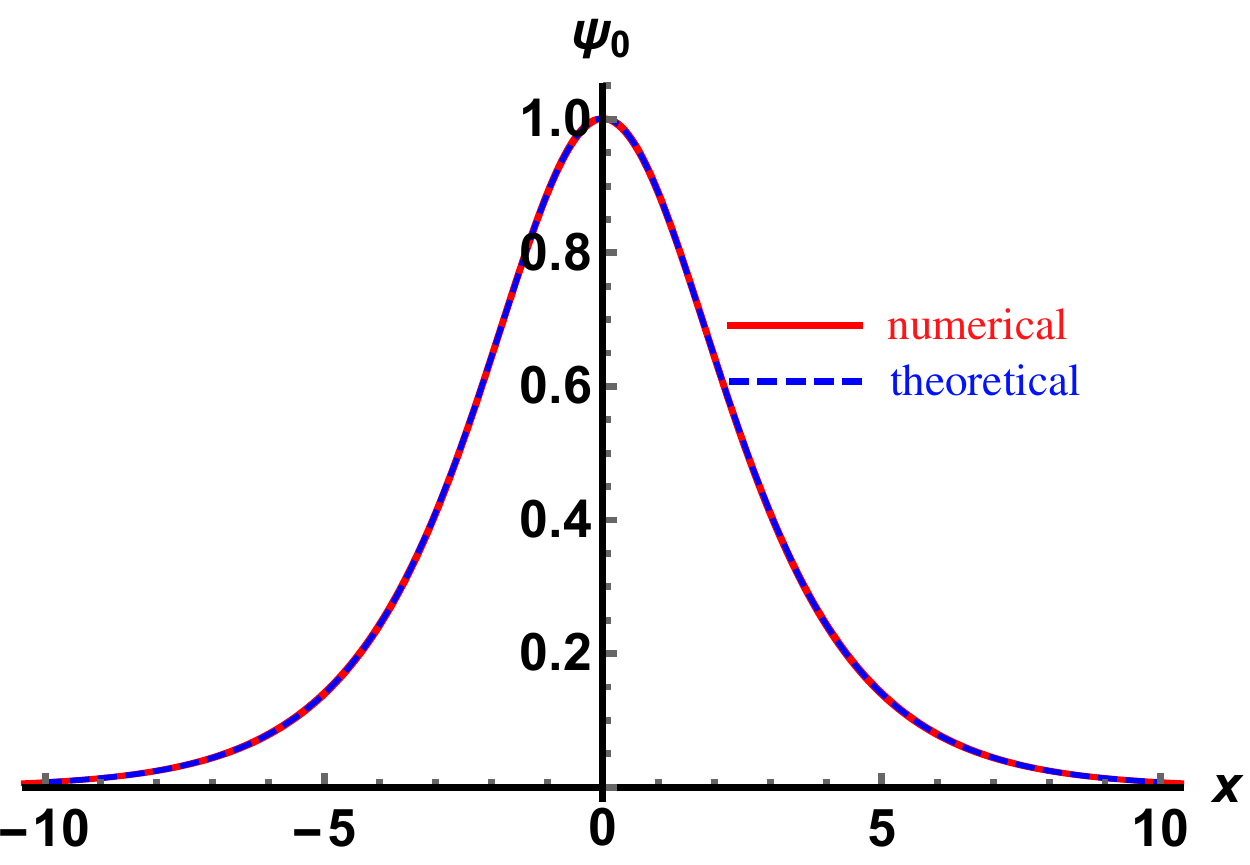}\label{fig:Sn3WaveFunction}}
    \subfigure[$n=3$, topological sector ($\frac{1}{3}, 1$)]
{\includegraphics[width=0.48\textwidth]{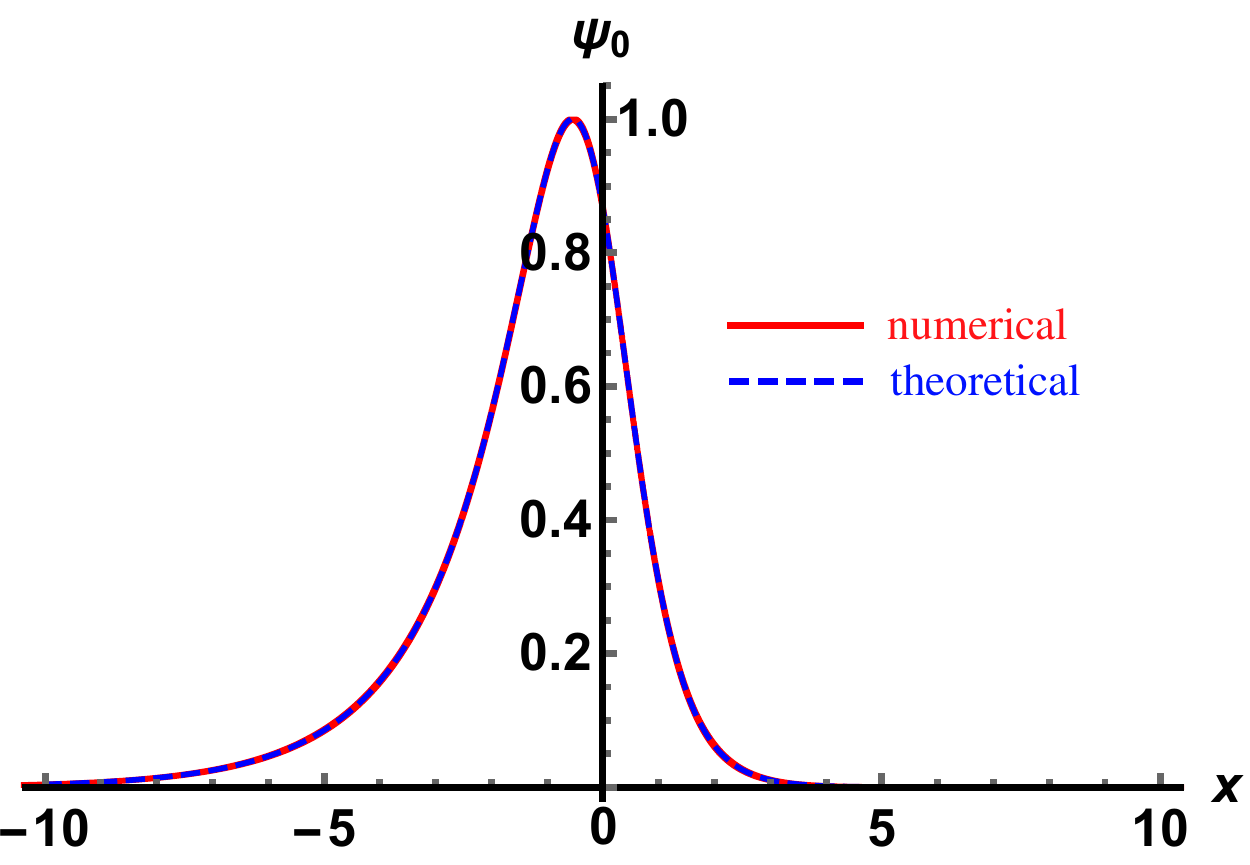}\label{fig:ASn3WaveFunction}}
  \\
  \caption{The zero mode wave functions obtained {\it numerically} from Eq.~\eqref{eq:stat_Schr} --- red solid lines, and {\it theoretical} functions $\displaystyle\frac{d\varphi_{\rm K}^{}}{d x}$ --- blue dashed lines. For convenience, all the functions are normalized to unity at the maximum points.}
  \label{fig:wavefunctions}
\end{center}
\end{figure}

Moreover, in all cases we have obtained eigenfunctions associated with the eigenvalues $\omega_0^2$. As it should be, they are nodeless wave functions corresponding to the ground states of the discrete spectrum. Up to normalization coefficients, the found eigenfunctions coincide with the derivatives of the kinks $\displaystyle\frac{d\varphi_{\rm K}^{}}{d x}$, as it should be \cite[Eq.~(2.19)]{Bazeia.EPJC.2018}, see Fig.~\ref{fig:wavefunctions}.

\section{Conclusion}
\label{sec:conclusion}

We have considered a field-theoretic model with a real scalar field in the $(1+1)$-dimensional space-time with field self-interaction (model potential) in the form of the eighth degree polynomial \eqref{eq:potential} with four degenerate minima. We were interested in the possibility of obtaining kink-type solutions in explicit form $\varphi=\varphi(x)$. Despite being in demand, kinks of the $\varphi^8$ model were still known only in an implicit form $x=x(\varphi)$ until now. This situation significantly limited the study of their properties, especially analytically.

We have shown that in the case of a ratio of constants $b/a=n$ equal to positive integers, in order to obtain explicit formulas for kinks, it is necessary to solve an algebraic equation of degree $n+1$. As an example, we have considered cases of $n=2$ and 3 and obtained analytical formulas for kinks in all topological sectors of the model. For $n=3$, the expressions for kinks look rather cumbersome; nevertheless, this is a significant step forward in the study of topological solitons of the $\varphi^8$ model.

Further, using the obtained formulas for kinks, we have calculated the kinks' masses, which surely coincide with the values obtained using the superpotential. The point is that the topological solitons under consideration are BPS-saturated static configurations that have the smallest possible energy in their topological sectors. This energy can be easily found using superpotential.

Besides that, using explicit formulas for kinks, we investigated the excitation spectra of all kinks at $n=2$ and $n=3$. A thorough search of levels in the discrete part of the spectrum of the eigenvalue problem \eqref{eq:stat_Schr} has shown the presence of only zero levels with $\omega_0^{}\approx 0$. This means that all the kinks considered in this paper have only translational modes.

Emphasize that this paper does not claim to be an exhaustive study of kinks of the $\varphi^8$ model. As was mentioned above, the potential of the $\varphi^8$ model can be written in different forms \cite{Khare.PRE.2014,Lohe.PRD.1979,Gani.JHEP.2015,Christov.PRD.2019,Belendryasova.CNSNS.2019,Christov.PRL.2019}. In Ref.~\cite{Christov.PRD.2019} it was shown that in some cases kinks with power-law tails may exist. In this our paper we focused on obtaining and studying kink-type solutions {\it in the explicit form} for a specific kind of the $\varphi^8$ model, which admits the existence of kinks with exponential tails. Of course, in all other variants of the $\varphi^8$ model, the option of obtaining of kink-type solutions numerically always remains available.

In conclusion, we would like to briefly mention several issues that have not been addressed in this paper but, in our opinion, could be of interest for future studies.
\begin{itemize}
    \item Searching for explicit kink solutions in models with polynomial potentials of higher degrees, e.g., $\varphi^{10}$, $\varphi^{12}$, etc., could become a natural extension of this study.
    \item A study of the general properties of algebraic equations that determine explicit kink solutions in the case of arbitrary $n$, as well as a study of the case $n\gg 1$ are of interest.
    \item Finally, another direction of current interest concerns using of the found explicit kink solutions for studying kink-antikink scattering. In particular, despite the absence of vibrational modes in the kink's excitation spectrum, in the case of collisions of asymmetric kinks one can expect the appearance of resonance phenomena due to resonant energy exchange. One of the vibrational modes in the collective ``quantum-mechanical'' potential of the system ``kink+antikink'' can play the role of an accumulating localized mode (such a mechanism was studied in Refs.~\cite{Dorey.PRL.2011,Belendryasova.CNSNS.2019}). In the context of the kink-antikink interactions, the forces between kink and antikink can also be estimated using the obtained asymptotics of the kinks (via the so-called Manton's method \cite{Manton.NPB.1979}).
\end{itemize}

\section*{Acknowledgments}


We are grateful to the PRD referees for their valuable and motivating comments and criticism that helped us to improve presentation of this our study.

The work of the MEPhI group was supported by the MEPhI Academic Excellence Project (Contract No.\ 02.a03.21.0005, 27.08.2013). V.A.G.\ also acknowledges the support of the Russian Foundation for Basic Research (RFBR) under Grant No.\ 19-02-00971. A.M.M.\ thanks the Islamic Azad University, Quchan Branch, Iran (IAUQ) for their financial support under the Grant.



\vspace{10mm}

\hrule

\vspace{20mm}

\begin{figure}[h!]
\centering
\includegraphics[width=0.2\textwidth]{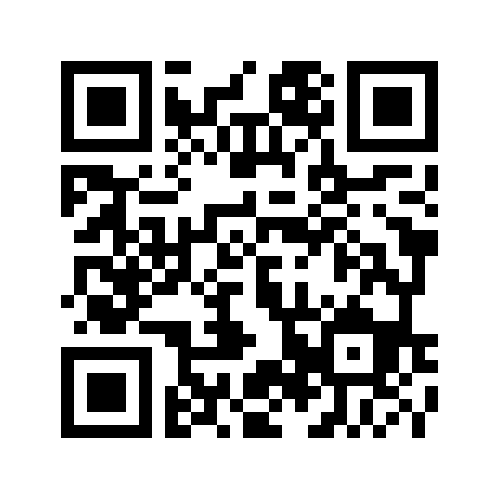}
\hspace{10mm}
\includegraphics[width=0.2\textwidth]{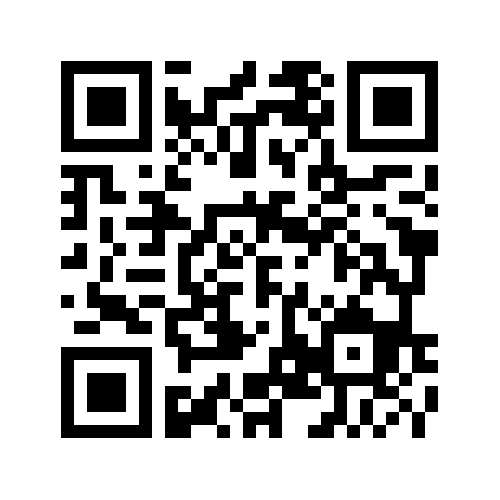}
\hspace{10mm}
\includegraphics[width=0.2\textwidth]{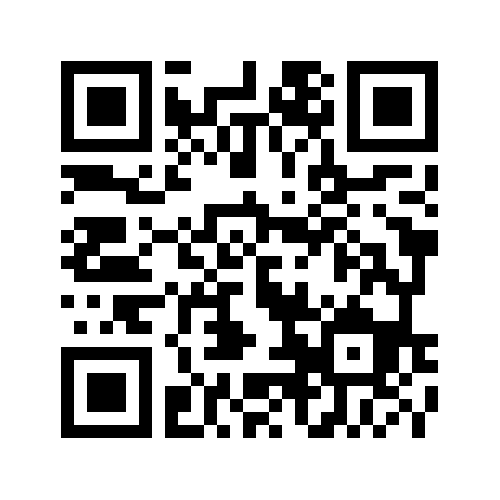}
\end{figure}

\end{document}